\documentclass[letterpaper, 10 pt, conference]{ieeeconf}  
\usepackage{graphicx}
\usepackage{amsmath,amsfonts,mathrsfs,amssymb,dsfont}
\usepackage{booktabs}
\usepackage{mathtools}
\usepackage{ stmaryrd }
\usepackage{cite}
\usepackage{sidecap}
\sidecaptionvpos{figure}{l}
\IEEEoverridecommandlockouts                              
   \usepackage{xcolor}                                                       
\usepackage{mathabx}

\usepackage{graphicx}

\def\moyenne#1{\langle #1\rangle}

\title{\LARGE \bf
Stochastic Analysis Of An Incoherent Feedforward Genetic Motif
}

\author{Thierry Platini$^{1}$, Mohammad Soltani$^{2}$, Abhyudai Singh$^{3}$
\thanks{$^{1}$T. Platini is with the Applied Mathematic Research Center (AMRC), Coventry University, UK.
        {\tt\small thierry.platini@coventry.ac.uk}}%
\thanks{$^{2}$M. Soltani is with the Department of Electrical and Computer Engineering, University of Delaware, Newark, DE USA 19716.
{\tt\small msoltani@udel.edu}}%
\thanks{$^{3}$A. Singh is with the the Department of Electrical and Computer Engineering, Biomedical Engineering, Mathematical Sciences, Center for Bioinformatics and Computational Biology, University of Delaware, Newark, DE USA 19716.
{\tt\small absingh@udel.edu}}}

\begin{document}

\maketitle
\thispagestyle{empty}
\pagestyle{empty}

\begin{abstract}
Gene products (RNAs, proteins) often occur at low molecular counts inside individual cells, and hence are subject to considerable random fluctuations (noise) in copy number over time.  Not surprisingly, cells encode diverse regulatory mechanisms to buffer noise. One such mechanism is the incoherent feedforward circuit. We analyze a simplistic version of this circuit, where an upstream regulator $X$ affects both the production and degradation of a protein $Y$. Thus, any random increase in $X$'s copy numbers would increase both production and degradation, keeping  $Y$ levels unchanged.  To study its stochastic dynamics, we formulate this network into a mathematical model using the Chemical Master Equation formulation. We prove that if the functional dependence of $Y$'s production and degradation on $X$ is similar, then the steady-distribution of $Y$'s copy numbers is independent of $X$. To investigate how fluctuations in $Y$ propagate downstream, a protein $Z$ whose production rate only depend on $Y$ is introduced. Intriguingly, results show that the extent of noise in $Z$ increases with noise in $X$, in spite of the fact that the magnitude of noise in $Y$ is invariant of $X$. Such counter intuitive results arise because $X$ enhances the time-scale of fluctuations in $Y$, which amplifies fluctuations in downstream processes. In summary, while feedforward systems can buffer a protein from noise in its upstream regulators, noise can propagate downstream due to changes in the time-scale of fluctuations.
\end{abstract}
\section{Introduction}
The inherent probabilistic nature of biochemical reactions and low copy numbers of molecules involved, results in significant random fluctuations (noise) in mRNA/protein levels inside individual cells \cite{rao08,bkc03,rao05,keb05,Eldar:2010kk,Neuert01022013,Chalancon2012221,Magklara:2014iq,jbp14}. These fluctuations are an unavoidable aspect of life at the single-cell level. Noise can be problematic for essential proteins whose levels have to be tightly maintained within certain bounds \cite{lps07,fhg04,leh08}, and various diseased states have been attributed to elevated noise in the expression of certain genes \cite{ksv02,cgt98,bhr06,bch09}. Interestingly, this inherent variation in gene product levels is sometimes exploited for driving genetically identical cells to different fates \cite{arm98,lod08,Balazsi:2014bv,Norman:kq,St-Pierre30122008,kys12}, as is the case for many stem cells \cite{chm08,Abranches01072014,Torres-Padilla01062014} and pathogenic human viruses \cite{wbt05,wds08,tps09,siw09}.

Given that stochasticity in protein levels can have significant effects on biological function and phenotype, cells actively use different regulatory mechanisms to minimize noise. Much prior experimental/computational work on noise buffering has primarily focused on negative feedback systems, where a protein controls its own transcription/translation/degradation  \cite{sak06,sih09c,lvp10,swa04,bhs08,sav74,bes00,nam09,dmk06,tho01,sih09b,dah04,sih07ny,hkr11}. Here we focus on feedforward systems, where a downstream regulator affects the expression of a protein using two different paths. More specifically, we study the \emph {incoherent feedforward loop}, where the paths have antagonistic affects \cite{bxg11,alo07,whs03}. Such incoherent feedforward regulation has been shown to be an important motif in gene regulatory networks \cite{alo07,whs03,tzo07}.

The schematic of the overall network is illustrated in Fig. 1 and consists of three species: an upstream regulator $X$, protein $Y$, and a downstream product $Z$ that is activated by $Y$. In the model under consideration, $X$ enhances both the production and degradation of $Y$,
creating an incoherent feedforward circuit. In the stochastic formulation of this network, each specie is assumed to be produced in random bursts \cite{shs08,pau05,sim13}. 
More specifically, bursts for the creation of $X$ arrive at exponentially distributed time intervals with rate $k_X$. Each burst results in the production of $n_X$ molecules of $X$, where $n_X$ is geometrically distributed random variable. $X$ is assumed to degrade at a constant rate $\gamma_X$. Finally, we denote by $x(t)$, the stochastic process representing the population count of $X$ in a single cell. The same nomenclature applies for $Y$ and $Z$.
\begin{figure}[thpb]
\centering
  \includegraphics[width=0.9\linewidth]{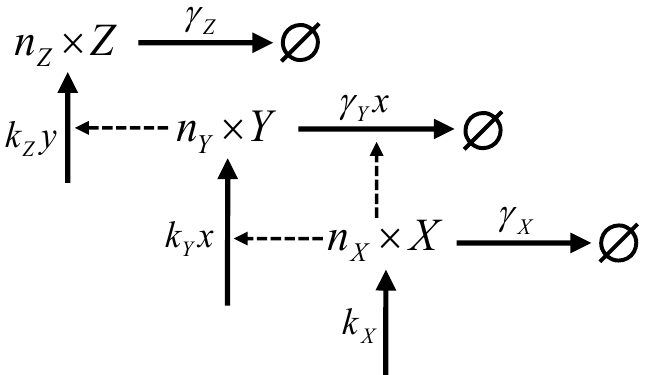}\\
  \caption{{\bf Schematic figure of the model under consideration.} 
$X$ affects both the production and degradation of Y, which itself activates production of downstream product $Z$. The creation and degradation rates of molecules $X$, $Y$ and $Z$ are denotes $k_X$, $\gamma_X$, $k_Y$, $\gamma_Y$ and $k_Z$,  $\gamma_Z$. Each creation event generates a burst, of size $n_{\bf j}$, characterized by a geometrical distribution $g(n_{\bf j}|\bar b_{\bf j})$ with mean ${\bar b_{\bf j}}$ (for ${\bf j}=X,Y,Z$).
  }  $\hphantom{.}$\\
        \label{figmodel1}
\end{figure}

Our goal is to use this model to study how random fluctuations in the levels of $X$ propagate to $Y$ and $Z$. Results show that if the functional dependence of $Y$'s production and degradation on $X$ is similar, then the steady-distribution of $Y$'s copy numbers is independent of $X$. Thus, the feedforward regulation completely buffers $Y$ from random fluctuations in the upstream regulator. Interestingly, fluctuations in $X$ enhance the time-scale of fluctuation in $Y$, as quantified by the steady-state autocorrelation function. This implies that fluctuations in $X$ make fluctuations in $Y$ more permanent while keeping their magnitude unchanged, leading to an amplified noise in the downstream product $Z$.

The paper is organized as follows. In section \ref{mod1}, we present a stochastic model for the expression of protein $Y$, with constant production and degradation rates.  In section \ref{mod2}, we consider the effect of the upstream regulator $X$ on $Y$'s production process. In section \ref{mod3}, $X$ is assumed to affect both production and degradation processes of $Y$, creating a feedforward system. The autocorrelation function of $Y$ is derived in section  \ref{mod3.1}. Finally, in section \ref{mod3.2} we quantify the noise in the downstream product $Z$.

\section{Single protein model with constant rates}\label{mod1}
We start by considering the model (summarised by table \ref{Table_mod1}) describing the dynamics of the number of molecules $Y$, with constant production and degradation rates.
\begin{table}[h!]
\centering
   \begin{tabular}{c c c}
   \hline
     Event	& Reset & Transition rates  \\  \hline	\hline   &\\
    burst of $n_Y$ $Y$ molecules & $y\rightarrow y+n_Y$ & $k_Yg(n_Y|{\bar b_Y})$ \\ \\
    degradation & $y\rightarrow y-1$ & $y\gamma_Y$ \\	\hline
	\end{tabular}
  \caption{Transitions and associated rates for the single protein model with constant rates}
  \label{Table_mod1}
\end{table}
\\
We write $k_Y\delta t$ the probability of a burst occurring in a time $\delta t$. Each burst is drawn from a geometric distribution:
\begin{eqnarray}
g(n_Y|{\bar b_Y})={(\bar b_Y)^{n_Y}}/{(1+{\bar b_Y})^{{n_Y}+1}},
\end{eqnarray} 
with mean ${\bar b_Y}=\sum_{{n_Y}}{n_Y}g({n_Y}|{\bar b_Y})$. 
In a time interval $\delta t$, the probability of occurrence of a burst of size ${n_Y}$  is therefore given by $k_Yg({n_Y}|{\bar b_Y})\delta t$. In addition, we denote by $\gamma_Y$ the degradation rate so that the probability of the transition, from a state with $y$ molecules to a state with $y-1$ molecules, in a time $\delta t$, is given by $y\gamma_Y \delta t$. It is well known that the probability $P_{y}(t)$ to measure $y$ molecules at time $t$, obeys the master equation \cite{Gillespie76,mcq67,wil11}
\begin{eqnarray}\label{Master_1}
\frac{d P_y(t)}{dt}&=&k_Y\left[\sum_{n_Y=0}^yg(n_Y|{\bar b_Y})P_{y-n_Y}(t)-P_y(t)\right]\nonumber\\
&+&\gamma_Y\Big[(y+1)P_{y+1}(t)- y P_y(t)\Big]. 
\end{eqnarray}
The latter equation gives a full description of the stochastic process under consideration. It is, however, common practice to express such problem in term of the generating function defined as $G(r,t)=\sum_yP_y(t)r^y$. Derived from \eqref{Master_1} the equation for the generating function is:
\begin{eqnarray}\label{masterG}
\frac{dG}{dt}=k_Y(\tilde g-1)G-\gamma_Y(r-1)\partial_rG,
\end{eqnarray}
where $\tilde g$ is the generating function of the distribution $g(n_Y|{\bar b_Y})$ and given by 
\begin{eqnarray}
\tilde g(r|{\bar b_Y})=\sum_{n_Y=0}^\infty r^{n_Y}g(n_Y|{\bar b_Y})=\frac{1}{1+{\bar b_Y}(1-r)}.
\end{eqnarray}
Equation \eqref{masterG} offers an easy path towards the solution of our problem. In particular, in the limit $t\rightarrow\infty$, we find 
\begin{eqnarray}
G(r)=\frac{1}{\left[1+{\bar b_Y}(1-r)\right]^{k_Y/\gamma_Y}}.
\end{eqnarray}
At this stage the inverse transfomation, 
\begin{eqnarray}
P(y=n)=\lim_{r\rightarrow0}\frac{1}{n!}\frac{d^nG(r)}{dr^n},
\end{eqnarray}
can be used to access the stationary probability distribution which, in this case, is a negative binomial distribution:
\begin{eqnarray}
P(y=n)&=&\frac{1}{n!}
\left(\frac{\bar b_Y}{1+\bar b_Y}\right)^n\left(\frac{1}{1+\bar b_Y}\right)^{\frac{k_Y}{\gamma_Y}}\\
&\times&
\prod_{j=0}^{n-1}\left(j+\frac{k_Y}{\gamma_Y}\right).\nonumber
\end{eqnarray}
One can directly access first and second order moments using
\begin{eqnarray}
\moyenne{y}=\left.\frac{dG(r)}{dr}\right|_{r\rightarrow1},\text{ and } \moyenne{y(y-1)}=\left.\frac{d^2G(r)}{dr^2}\right|_{r\rightarrow1},
\end{eqnarray}
which leads to the mean number
\begin{eqnarray}\label{simpleresultsm}
\moyenne{y}=\frac{k_Y{\bar b_Y}}{\gamma_Y}.
\end{eqnarray}
We will use the coefficient of variation $CV_Y^2$ as metric for quantifying noise. It is defined by
\begin{eqnarray}\label{simpleresults}
CV_Y^2= \frac{\moyenne{y^2}-\moyenne{y}^2}{\moyenne{y}^2},\nonumber
\end{eqnarray}
and given by
\begin{eqnarray}\label{simpleresults}
CV_Y^2=\frac{1+{\bar b_Y}}{\moyenne{y}}.
\end{eqnarray}
Unless stated otherwise $\moyenne{y}$ denotes the average in the stationary state. We will be explicitly using $\moyenne{y(t)}$ to refer to the average number at intermediate times. At this point, the reader may want to consider a similar problem for a non-bursty production (each production event generating exactly one molecule). To proceed, the reader may simply replace $k_Y$ by $k_Y/{\bar b_Y}$ and take the limit ${\bar b_Y}\rightarrow0$. This transformation comes from the need to reduce the term $g(r|{\bar b_Y})-1$, in \eqref{masterG}, into $y-1$. Under this transformation we recover the Poisson distribution 
\begin{eqnarray}
G(r)\rightarrow \exp\left[{\frac{k_Y}{\gamma_Y}(r-1)}\right],
\end{eqnarray}
characterized by the mean and coefficient of variation
\begin{eqnarray}
\moyenne{y}\rightarrow\frac{k_Y}{\gamma_Y},\text{ and }CV_Y^2\rightarrow\frac{1}{\moyenne{y}}.
\end{eqnarray}

\section{Regulation of the creation process}\label{mod2}
We now focus our attention on a variation of the model presented in section  \ref{mod1} for which molecule production is regulated by an upstream process. Here the bursty creation process of $Y$ is governed by another dynamical process. A new random variable $x$ is introduced describing the number of molecules of type $X$. The extra dynamical variable $x$ now appears explicitly in the bursty production rate which becomes $xk_Y$. We choose to consider $X$ as governed by a bursty creation process with single degradation. We write $k_X$ and $\gamma_X$ the creation and degradation rates. Each burst of $X$ is distributed by $g({n_X}|{\bar b_X})$ where ${\bar b_X}$ denotes the mean burst size. All transition rates are summarized in table \ref{table_mod2}. We choose not to write the full master equation associated to the evolution of the distribution $P_{x,y}(t)$. We however give the key steps leading to the moment equations. To proceed the reader may derive a generalized moment equation \cite{sih10,sih10a}
\begin{eqnarray}
\frac{d\moyenne{x^\sigma y^\eta}}{dt}&=&
k_X \moyenne{\left[(x+n_X)^\sigma-x^\sigma\right] y^\eta}\\
&+&
\gamma_X
 \moyenne{x\left[(x-1)^\sigma -x^\sigma\right] y^\eta }\nonumber\\
&+&
k_Y\moyenne{x^{\sigma+1}\left[ (y+n_Y)^\eta - y^\eta\right]}
\nonumber\\
&+&
\gamma_Y
\moyenne{x^\sigma \left[(y-1)^\eta-y^\eta\right] y},
\nonumber
\end{eqnarray}
for $\sigma$ and $\eta$ integers. The latter equation leads to the first order moments
\begin{eqnarray}
\frac{d\moyenne{x(t)}}{dt}&=&k_X {\bar b_X}-\gamma_X\moyenne{x(t)},\\
\frac{d\moyenne{y(t)}}{dt}&=&k_Y {\bar b_Y}\moyenne{x(t)}-\gamma_Y\moyenne{y(t)},
\end{eqnarray}
as well as second order moments
\begin{eqnarray}
\frac{d\moyenne{x^2(t)}}{dt}&=&k_X {\bar b_X}[2\moyenne{x(t)}+2{\bar b_X}+1]\\
&+&\gamma_X[\moyenne{x(t)}-2\moyenne{x^2(t)}],\nonumber\\
\frac{d\moyenne{y^2(t)}}{dt}&=&k_Y{\bar b_Y}\left[
2\moyenne{xy(t)}+(2{\bar b_Y}+1)\moyenne{x(t)}
\right]\\
&+&\gamma_Y\left[\moyenne{y(t)}-2\moyenne{y^2(t)}\right],
\nonumber\\
\frac{d\moyenne{xy(t)}}{dt}&=&k_X {\bar b_X}\moyenne{y(t)}+k_Y{\bar b_Y}\moyenne{x^2(t)}\\
&-&(\gamma_X+\gamma_Y)\moyenne{xy(t)}.\nonumber
\end{eqnarray}
\begin{table}[h!]
\centering
   \begin{tabular}{c c c}
   \hline
     Event	& Reset & Transition rates  \\  \hline	\hline  & \\
    {burst of $n_X$ $X$ molecules}  & {$x\rightarrow x+n_X$} & {$k_X g(n_X|{\bar b_X})$} \\ \\
    {$X$-degradation} & {$x\rightarrow x-1$} & {x$\gamma_X$} \\	\\
    burst of $n_Y$ $Y$ molecules & $y\rightarrow y+n_Y$ & $ {x} k_Y g(n_Y|{\bar b_Y})$ \\ \\
    $Y$-degradation & $y\rightarrow y-1$ & $y \gamma_Y$ \\	\hline
	\end{tabular}
  \caption{Transitions and associated rates for a model with regulated creation process.}
  \label{table_mod2}
\end{table}
\vspace{-5mm}

The previous set of equations being closed one can easily show that the stationary state is characterized by the mean numbers
\begin{eqnarray}\label{eqmeanmod2}
\moyenne{x}=\frac{k_X {\bar b_X}}{\gamma_X} \text{ and }
\moyenne{y}=\frac{k_Y {\bar b_Y}}{\gamma_Y}\moyenne{x},
\end{eqnarray}
with the following coefficients of variation
\begin{eqnarray}
CV_X^2&=&\frac{1+{\bar b_X}}{\moyenne{x}}, \\
CV_Y^2&=&\frac{1+{\bar b_Y}}{\moyenne{y}}+\frac{\gamma_Y}{\gamma_X+\gamma_Y}\frac{1+{\bar b_X}}{\moyenne{x}}.
\label{EQ_21}
\end{eqnarray}
Note that $CV_Y^2$ is the sum of two contributions. The first term represents the noise in the single protein model with constant rates. The second term is the noise contribution from upstream regulation. We note  that both $\moyenne{y}$ and $CV_Y^2$ are dependent on the upstream dynamics (dependence in $k_X$, $b_X$ and $\gamma_X$). The dependence in the $X$ dynamics will however vanished in the next section when considering regulated production and degradation.

\section{Incoherent feedforward circuit}\label{mod3}
To move one step forward we choose to consider a model where both the production and degradation are affected by the dynamics of $X$, therefore defining a feedforward motif. We define $xk_Y$ and $x\gamma_Y$ as the new creation and degradation rates. All transition rates are summarized in table \ref{table_mod3} and illustrated on Fig. \ref{figmodel1}. 
\begin{table}[!h]
\centering
   \begin{tabular}{c c c}
   \hline
     Event	& Reset & Transition rates  \\  \hline	\hline & \\
    burst of $n_X$ $X$ molecules  & $x\rightarrow x+n_X$ & $k_X g(n_X|{\bar b_X})$ \\ \\
    $X$-degradation & $x\rightarrow x-1$ & $x\gamma_X $ \\	\\
    burst of $n_Y$ $Y$ molecules & $y\rightarrow y+n_Y$ & $xk_Yg(n_Y|{\bar b_Y})$ \\ \\
    $Y$-degradation & $y\rightarrow y-1$ & $xy\gamma_Y$ \\	\hline
	\end{tabular}
  \caption{Transitions and associated rates for an incoherent feedforward circuit.}
  \label{table_mod3}
\end{table}
\vspace{-5mm}
\\
The probability $P_{x,y}(t)$ is governed by the master equation
\begin{eqnarray}\label{eqPXp}
\frac{d P_{x,y}(t)}{dt}&=&
k_X \left[\sum_{n_X=0}^xg(n_X|{\bar b_X})P_{x-n,y}-P_{x,y}\right]\\
&+&\gamma_X \left[(x+1)P_{x+1,y}-xP_{x,y}\right]\nonumber\\
&+&xk_Y\left[\sum_{n_Y=0}^yg(n_Y|{\bar b_Y})P_{x,y-n}-P_{x,y}\right]\nonumber\\
&+&x\gamma_Y\Big[(y+1)P_{x,y+1}-y P_{x,y}\Big].\nonumber
\end{eqnarray}
It is then important to note that, in the stationary state, writing $P_{x,y}=Q_{x}R_{y}$ allows to split \eqref{eqPXp} in two:
\begin{eqnarray}
\label{eqA}
k_X\sum_{n_X=0}^xg(n_X|{\bar b_X})Q_{x-n}+\gamma_X (x+1)Q_{x+1}&\\
=(k_X+\gamma_X x)Q_{x},&\nonumber\\
\label{eqB}
k_Y\sum_{n_Y=0}^y g(n_Y|{\bar b_Y})R_{y-n}+\gamma_Y(y+1)R_{y+1}&\\
=(k_Y+\gamma_Y y)R_{y}.&\nonumber
\end{eqnarray}
Note that \eqref{eqA} and \eqref{eqB} have exactly the same form but more importantly are independent. It follows that the $x$ and $y$ variables are uncorrelated. The average $\moyenne{y}$ and all other moments are independents of $k_X$, $\gamma_X$ and ${\bar b_X}$. It is important to mention that an identical derivation can be repeated in a much more general scenario: First by generalizing this result for any production and degradation rates of the form $k_Y(x)=k_Yf(x)$ and $\gamma_Y(x)=\gamma_Y f(x)$ (for an arbitrary function $f$). Secondly by relaxing constrains on the dynamics of $X$ and writing $W^x_{x'}$ as the transition rate associated to $x\rightarrow x'$ (with the restriction that $W^x_{x'}$ is independent of $y$). Once again, a similar derivation will show that the mean number of $Y$ molecules and all moments are independent of the upstream process associated to $X$.

As a consequence, when looking at the stationary distribution of $x$ only, one would not be able to distinguish the model with feedfoward motif (regulated creation and degradation) from the single protein model (with no input noise at all). In a sense, the $x$ variable and its dynamics are ``hidden" in the stationary state. However, a signature of this ``hidden variable" may be observed someplace else. Indication that the process is or not governed by a ``hidden" dynamics could be found in dynamical data. Since the equality $P_{x,y}=Q_xR_y$ holds true in the stationary state only, the analysis of transient regime should give evidences of the upstream process. For example, one could study quantities such as relaxation time and autocorrelations, which would, in principle, testify of the existence of $X$. Interestingly, another indication of the existence of an upstream noise is to be found in downstream production. In the next sections we look for signature of an upstream regulator in both autocorrelation function and downstream processes.
\section{Effect of Feedforward regulation on autocorrelation time}\label{mod3.1}
In the following we present an analytical study of the autocorrelation function. We first start with a presentation of the method used, considering the single protein model with constant production rates  (summarized in table \ref{Table_mod1}). The $Y$ autocorrelation function for the model presented in table \ref{table_mod3} and illustrated in Fig. \ref{figmodel1} is however unknown and its calculation appears extremely challenging. We therefore consider a feeforward model regulated by a binary process as illustrated in Fig. \ref{figmodel4} and summarized in table \ref{table_mod_XXXX}. 
In the stationary state, it is common to study the normalized autocorrelation function defined by
\begin{equation}\label{auto corr0}
R(t):=\lim_{s\rightarrow\infty}\frac{\langle y(t+s)y(s) \rangle -\moyenne{y}^2}
{\moyenne{y^2} - \moyenne{y}^2}.
\end{equation}
To progress further we use the relation
\begin{equation}
        \begin{aligned}
& \langle y(t+s)y(s) \rangle =\langle y(s) \langle  y(t+s) \vert y(s)  \rangle \rangle,
  \label{ts0}
        \end{aligned}
\end{equation}
where $\langle  y(t+s) \vert  y(s)  \rangle$ is the expected number of molecules at time $t+s$ given $y(s)$ \cite{singh_consequences_2012,moh15b}. Using theorem 1 of \cite{hsi04}, we see that the time derivative of the expected value of any function $\varphi(y)$ is given by
\begin{equation}
\begin{aligned}
 \frac{d\langle \varphi(y ) \rangle}{dt}= \left \langle  \sum_{Events}  \Delta \varphi(y)  \times f(y )  \right \rangle, \label{dynnf}
  \end{aligned}
\end{equation}
where $\Delta \varphi$ is change in function $ \varphi$ when an event occurs and $f(y)$ denotes the transition rates of events and shows how often an event happens.
\subsection{Single protein model with constant rates} Considering the model with single protein, equation \eqref{dynnf} for $\varphi(y)=y$ gives 
\begin{equation}
\frac{d\moyenne{y(t)}}{dt}=k_Y{\bar b_Y}-\gamma_Y\moyenne{y(t)},
\end{equation}
and leads to the stationary values given in \eqref{simpleresultsm}. It follows that the mean count at time $t$ knowing $y(s)$ is given by
\begin{equation}\label{cdm}
        \begin{aligned}
&\langle y(t+s) \vert y(s)  \rangle= {\langle y \rangle} + \left[ y(s)- {\langle y \rangle} \right] e^{-\gamma_Y  t}. \\
        \end{aligned}
\end{equation}
Using \eqref{ts0} together with \eqref{cdm} leads to
\begin{equation}
        \begin{aligned}
&\langle y(t+s)y(s) \rangle = {\langle y \rangle}^2  + \left[ {\langle y^2 \rangle} - {\langle y \rangle}^2  \right] e^{-\gamma_Y t}.
  \label{ts2}
        \end{aligned}
\end{equation}
By substituting \eqref{ts2} in \eqref{auto corr0} we obtain the autocorrelation function which appears to be completely determined by the  degradation rate:
\begin{align}\label{auto corr00}
& R(t)=\exp\left({-\gamma_Y t}\right).
\end{align} 
\vspace{-5mm}
\begin{table}[h!]
\centering
   \begin{tabular}{c c c}
   \hline
     Event	& Reset & Transition rates  \\  \hline \hline
    Switch activation & $x\rightarrow x+1$ & $(1-x)\alpha$ \\  \\
    Switch deactivation & $x\rightarrow x-1$ & $x\beta$ \\	\\
    burst of $n_Y$ $Y$ molecules & $y\rightarrow y+n_Y$ & $xk_Yg(n_Y|{\bar b_Y})$ \\ \\ 
    $Y$-degradation & $y\rightarrow y-1$ & $xy\gamma_Y$ \\	\hline
	\end{tabular}
  \caption{Transitions and associated rates for a model regulated by a biological switch.}
  \label{table_mod_XXXX}
\end{table}
\vspace{-5mm}
\subsection{A model regulated by a biological switch}
\begin{figure}[thpb]
\centering
  \includegraphics[width=0.9\linewidth]{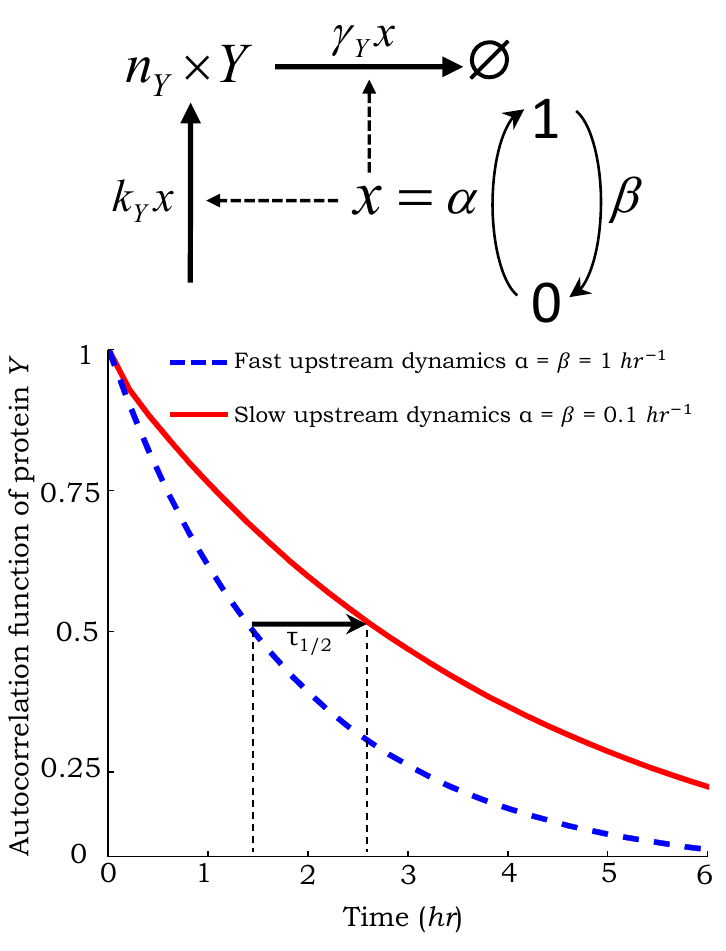}\\
\caption{\textbf{Autocorrelation in feedforward motif with biochemical switch.} \textit{Top}: Feedforward motif: both production and degradation rate of molecules $Y$ are dependent of upstream regulator $x$. Upstream process is restricted to the values $x=0$ and $x=1$. Its dynamics is governed by transition rates $ (1-x) \alpha$ and $x \beta $. \textit{Bottom}: $Y$-autocorrelation function as a function of the time. All the rates are normalized to protein decay rate, i.e., $\gamma_Y=1\ hr^{-1}$. Data are showing a shift of $\tau_{1/2}$ (time at which $R(t)$ reaches $50\% $ of its initial value) due to upstream regulator dynamics. }
\label{autocor}
\end{figure}

We now evaluate the autocorrelation function for a model where the upstream regulator is restricted to the values $x=0$ and $x=1$ (see Fig. \ref{autocor}). This model, should be regarded as a first step towards a more complex model. In fact, H. Pendar and collaborators \cite{ppk13} have shown that any birth-death process can be split into an infinite number of identical reduced models, each build as biological switches. In this picture, biological switches appears as the basic construction brick for more sophisticated models. We write $\alpha$ and $\beta$ the transition rates associated $x:0\rightarrow1$ and $x:1\rightarrow0$ and summarized in table \ref{table_mod_XXXX}. This chemical switch regulates both bursty production and degradation of molecules of type $Y$. For this particular model, we derive the moment equations
\begin{eqnarray}
\frac{d \langle x(t)\rangle}{dt} &=& \alpha - (\alpha + \beta) \langle x (t)\rangle,
\label{eqx}\\
\label{eqy}
\frac{d \langle y (t)\rangle}{dt} &=& k_Y {\bar b_Y} \langle x (t)\rangle   - \gamma_Y \langle x y  (t)\rangle,
\\
\frac{d   \langle x y (t)\rangle}{dt}&=& \alpha\langle y(t)  \rangle - (\alpha + \beta) \langle xy(t) \rangle
\\
&+& k_Y {\bar b_Y} \langle x (t)\rangle   - \gamma_Y \langle x y (t)\rangle.\label{eqxy}\nonumber
\label{ode}
\end{eqnarray}
Solving equations \eqref{eqx}-\eqref{eqxy} for initial conditions $x(s)$, $y(s)$ and $xy(s)$ result in
\begin{eqnarray}
& &\langle y(t+s)\vert x(s), y(s), xy(s) \rangle = {\langle y \rangle} \\
&+&  {\exp}\left[{-(\alpha+\beta+\gamma_Y)\frac{t}{2} }\right]  \nonumber \\
&\times& \frac{\sinh \left(\frac{t}{2} \sqrt{(\alpha+\beta+\gamma_Y )^2-4  \alpha \gamma_Y}\right)}{\sqrt{(\alpha+\beta+\gamma_Y)^2-4  \alpha \gamma_Y}} 
  \nonumber \\
&\times&((\alpha+\beta+\gamma_Y ) \left[ y(s)- {\langle y \rangle} \right]+2 \bar{b}_Y  \left[ x(s)- {\langle x \rangle} \right]  \nonumber \\
&-& 2 \gamma_Y \left[ xy(s)- {\langle xy \rangle} \right]) \nonumber\\
&+&{\exp}\left[{-(\alpha+\beta+\gamma_y)\frac{t}{2} }\right] \nonumber\\
&\times&\cosh \left(\frac{1}{2}t \sqrt{(\alpha+\beta+\gamma_Y)^2-4   \alpha \gamma_Y}  \right) \left[ y(s)- {\langle y \rangle} \right]
\nonumber 
\end{eqnarray}
Together with \eqref{ts0} the previous result leads to
\begin{eqnarray}
        \label{yys}
& &\langle y(t+s)y(s) \rangle = {\langle y \rangle}^2  \\
&+& (\alpha+\beta+\gamma_Y ) {\exp}\left[{- (\alpha+\beta+\gamma_Y)\frac{t}{2}}\right] \nonumber\\
&\times& \frac{\sinh \left(\frac{t}{2}\sqrt{(\alpha+\beta+\gamma_Y )^2-4  \alpha \gamma_Y}\right)}{\sqrt{(\alpha+\beta+\gamma_Y)^2-4  \alpha \gamma_Y}} \left( {\langle y^2 \rangle} - {\langle y \rangle}^2  \right) \nonumber\\
&+& \exp\left[{-(\alpha+\beta+\gamma_y) \frac{t}{2} }\right] \nonumber\\
&\times& \cosh \left(\frac{t}{2} \sqrt{(\alpha+\beta+\gamma_Y)^2-4   \alpha \gamma_Y}  \right) \left( {\langle y^2 \rangle} - {\langle y \rangle}^2  \right)\nonumber\\
&+&  {\exp}\left[{-(\alpha+\beta+\gamma_Y)\frac{t}{2} }\right].  \nonumber \\
&\times& \frac{\sinh \left(\frac{t}{2} \sqrt{(\alpha+\beta+\gamma_Y )^2-4  \alpha \gamma_Y}\right)}{\sqrt{(\alpha+\beta+\gamma_Y)^2-4  \alpha \gamma_Y}} \nonumber\\
&\times & \left( 2 \bar{b}_Y  \left({\langle xy \rangle} - {\langle x \rangle}{\langle y \rangle}  \right)-2\gamma_Y \left({\langle xy^2 \rangle} - {\langle xy \rangle}{\langle y \rangle}  \right) \right)\nonumber
\end{eqnarray}
In order to find $\langle y(t+s)y(s) \rangle $ we need to find expression of $\langle x y^2 \rangle$. Thus we add dynamics of $\langle y^2 \rangle$ and $\langle x y^2 \rangle$ to the set of moments dynamics presented in \eqref{eqx}-\eqref{eqxy}. 
Note that $x(t)$ is a Bernoulli random variable thus we have the following relations
\begin{equation}
\begin{aligned}
\langle x^q \rangle=\langle x\rangle,  \ \ \
\langle x^q y^b \rangle=\langle xy^b \rangle, \ \ \ q\in \{1,2,\ldots\}. 
\end{aligned}
\end{equation}
Using these characteristics, the moments dynamics of $\langle y^2 \rangle$ and $\langle x y^2 \rangle$ can be written as
\begin{eqnarray}
 \label{ode02}
\frac{d \langle y^2 (t)\rangle}{dt} &=& k_Y \bar{b_Y}(2\bar{b}_Y+1) \langle x (t)\rangle  +2 k_Y {\bar b_Y} \langle xy (t)\rangle \nonumber \\
&+& \gamma_Y \langle x y  (t)\rangle - 2 \gamma_Y \langle x y^2  (t)\rangle,\\
\frac{d \langle xy^2 (t)\rangle}{dt} &=& k_Y \bar{b}_Y (2\bar{b}_Y+1)  \langle x (t)\rangle  +2 k_Y {\bar b_Y} \langle xy (t)\rangle \nonumber\\
&+&  \gamma_Y \langle x y  (t)\rangle \nonumber - 2 \gamma_Y \langle x y^2  (t)\rangle \nonumber \\
&-&(\alpha + \beta) \langle x y ^2 (t)\rangle+\alpha \langle y^2 (t)\rangle.
\label{ode021}
\end{eqnarray}

Solving set of moments \eqref{eqx}-\eqref{eqxy}, \eqref{ode02}, and \eqref{ode021} in steady-state results in 
\begin{eqnarray}
&{\langle xy \rangle} - {\langle x \rangle}{\langle y \rangle}=\frac{\alpha  k_x \bar{b_Y} }{\gamma_Y (\alpha+\beta)} -\frac{ k_x \bar{b}_Y }{\gamma_Y} \frac{\alpha }{\alpha+\beta} =0, \label{xy}\\
&{\langle xy^2 \rangle} - {\langle xy \rangle}{\langle y \rangle}=\frac{\alpha }{\alpha+\beta} \frac{k_x \bar b_Y(\bar b_Y +1)}{ \gamma_Y}. \label{xyy}
\end{eqnarray}
Thus, putting equations \eqref{xy} and \eqref{xyy} in \eqref{yys} and using \eqref{auto corr0} lead to
\begin{eqnarray}
\label{eqRsumary}
R(t) & =& (\alpha+\beta+ \gamma_Y \frac{\beta-\alpha}{\alpha+\beta}) {\exp}\left[{-(\alpha+\beta+\gamma_Y) \frac{t}{2} }\right] \nonumber\\
&\times& \frac{\sinh \left(\frac{t}{2} \sqrt{(\alpha+\beta+\gamma_Y )^2-4  \alpha \gamma_Y}\right)}{\sqrt{(\alpha+\beta+\gamma_Y)^2-4  \alpha \gamma_Y}}  \\
&+&\exp\left[{- (\alpha+\beta+\gamma_Y)\frac{t}{2}}\right]Ê\nonumber\\
&\times& \cosh \left(\frac{t}{2} \sqrt{(\alpha+\beta+\gamma_Y)^2-4   \alpha \gamma_Y}  \right).\nonumber
\end{eqnarray}
\vspace{-5mm}
\\
The autocorrelation function for different values of $\alpha$ and $\beta$ is shown in Fig. \ref{autocor}. One should note that $R(t)$ is independent of the creation process and its parameters $k_Y$ and ${\bar b_Y}$. It is however strongly dependent on the upstream process. Note that when taking the limit $\beta\rightarrow0$ one recover the model with single protein for which $R(t)=\exp(-\gamma_Yt)$. It is particularly useful to define the ratio
\begin{eqnarray}
\label{eqRratio}
\Gamma(t)=\frac{R(t)}{R(t)|_{\beta\rightarrow0}},
\end{eqnarray}
for which we can show $\Gamma(t)\ge1$ $\forall t$. In other words; the feedforward circuit leads to a systematic increase of the autocorrelation function. The increase of the time scale of fluctuations is therefore expected to lead to a larger noise values in further downstream products. 
\section{Effect of regulation in further downstream products}\label{mod3.2}
\begin{figure}[thpb]
\centering
  \includegraphics[width=0.9\linewidth]{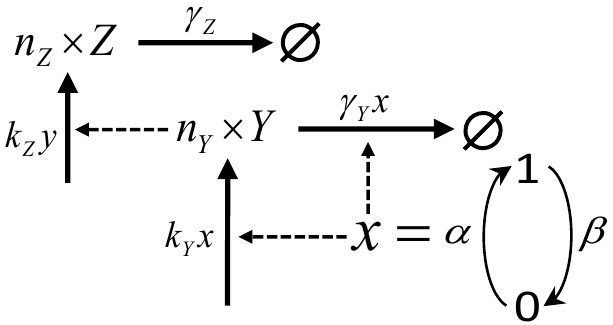}\\
  \caption{
{\bf Schematic figure of a feedforward circuit regulated by a biological switch.} $x$ is restricted to values $0$ and $1$ and governed by rates $\alpha$ and $\beta$. It affects both the production and degradation of Y, which itself activates production of downstream product $Z$. The creation and degradation rates of molecules $Y$ and $Z$ are denotes $k_Y$, $\gamma_Y$ and $k_Z$,  $\gamma_Z$. Each creation event generates a burst, of size $n_{\bf j}$, characterized by a geometrical distribution $g(n_{\bf j}|\bar b_{\bf j})$ with mean ${\bar b_{\bf j}}$ (for ${\bf j}=Y,Z$).
}
        \label{figmodel4}
\end{figure}

In this section we show that a signature of upstream input noise can be found in downstream products. We continue by considering the model, illustrated in Fig. \ref{figmodel4}, for which each molecule $Y$ can give birth to a burst of molecules $Z$. We write $k_Z$, $\bar b_Z$ and $\gamma_Z$ the associated creation rate, mean burst size and degradation rate. All transition rates are summarized in table \ref{table_mod4}.
\begin{table}[h!]
\centering
   \begin{tabular}{c c c}
   \hline
     Event	& Reset & Transition rates  \\  \hline	\hline
Switch activation & $x\rightarrow x+1$ & $(1-x)\alpha $\\ \\
Switch deactivation & $x\rightarrow x-1$ & $x\beta$\\	\\
    burst of $n_Y$ $Y$ molecules & $y\rightarrow y+n_Y$ & $xk_Yg(n_Y|{\bar b_Y})$ \\ \\
    $Y$-degradation & $y\rightarrow y-1$ & $xy\gamma_Y $ \\	\\
     burst of $n_Z$ $Z$ molecules  & $z\rightarrow z+n_Z$ & $ y k_Zg(n_Z|{\bar b_Z})$ \\ \\
    $Z$-degradation & $z\rightarrow z-1$ & $z\gamma_Z $ \\ \hline
	\end{tabular}
  \caption{Transitions and associated rates for a feedforward model regulated by a biological switch.}
  \label{table_mod4}
\end{table}
\vspace{-5mm}
\\
Once again, to spare the readers patience, we choose not to write the full master equation. The reader could however convinced himself that the dynamics of regulator $X$ should leave a trace in downstream production. To proceed one could verify that the probability $P_{x,y,z}$ can not be written as product of marginal probabilities $Q_x\times R_{y,z}$. The generalized moment equation for this model is
\begin{eqnarray}
\frac{d\moyenne{x^\sigma y^\eta z^\nu}}{dt}
&=&
(1-\delta_{\sigma,0})
\left[\alpha\moyenne{y^\eta z^\nu}-(\alpha+\beta)\moyenne{xy^\eta z^\nu}\right]
\nonumber\\
&+&
k_Y\moyenne{x^{\sigma+1}\left[ (y+n_Y)^\eta - y^\eta\right]z^\nu} \nonumber\\
&+&
\gamma_Y \moyenne{x^{\sigma+1} \left[(y-1)^\eta-y^\eta\right] yz^\nu}\nonumber\\
&+&
k_Z\moyenne{x^{\sigma}y^{\eta+1}\left[ (z+n_Z)^\nu - z^\nu\right]}\nonumber\\
&+&
\gamma_Z \moyenne{x^{\sigma} y^\eta \left[(z-1)^\nu-z^\nu\right] z},
\end{eqnarray}
for $\sigma$, $\eta$ and $\nu$ integers and where $\delta_{\sigma,0}=1$ for $\sigma=0$ and zero otherwise. From the latter equation, \eqref{eqx} and \eqref{eqy} can be derived together with
\begin{eqnarray}
\frac{d\moyenne{z(t)}}{dt}&=&k_Z\bar b_Y\moyenne{y(t)}-\gamma_Z \moyenne{z(t)}.
\end{eqnarray}
It follows that the stationary state is characterized by the mean numbers
\begin{eqnarray}
\moyenne{x}=\frac{\alpha}{\alpha+\beta}, \ \moyenne{y}=\frac{k_Y\bar b_Y}{\gamma_Y}, \  \moyenne{z}=\frac{k_Z\bar b_Z}{\gamma_Z}\moyenne{y}.
\end{eqnarray}
We should note that both mean numbers $\moyenne{y}$ and $\moyenne{z}$ show no dependence on upstream regulation dynamics. However, we see that second order moment $\moyenne{z^2}$ is a function of the correlation term $\moyenne{yz}$:
\begin{eqnarray}
\frac{d\langle z^2(t)\rangle}{dt}&=&2k_Z\bar b_Z\langle yz(t)\rangle
+k_Z(2\bar b_Z+1)\bar b_Z \langle y(t)\rangle \nonumber\\
&-&2\gamma_Z\langle z^2(t)\rangle
+\gamma_Z\langle z(t)\rangle.
\end{eqnarray}
In the stationary state, the latter equation leads to
\begin{eqnarray}
\moyenne{z^2}-\moyenne{z}^2=\moyenne{z}(1+{\bar b_Z})
+\frac{k_Z{\bar b_Z}}{\gamma_Z}\left(\moyenne{yz}-\moyenne{y}\moyenne{z}\right).
\end{eqnarray}
To move forward we derive the moment equation for $\moyenne{yz}$:
\begin{eqnarray}
\frac{d\moyenne{yz(t)}}{dt}&=&
k_Z{\bar b_Z}\moyenne{y^2(t)}+k_Y{\bar b_Y}\moyenne{xz(t)}\\
&-&
\gamma_Z\moyenne{yz(t)}-\gamma_Y\moyenne{xyz(t)}\nonumber,
\end{eqnarray}
which, in the stationary state, becomes
\begin{eqnarray}
\gamma_Z\moyenne{yz}+\gamma_Y\moyenne{xyz}=k_Z{\bar b_Z}\moyenne{y^2}+k_Y{\bar b_Y}\moyenne{xz}.
\end{eqnarray}
Along the same line, we derive equations for $d\moyenne{xz}/dt$ and $d\moyenne{xyz}/dt$. Taking the limit $t\rightarrow\infty$ leads to: 
\begin{eqnarray}
(\alpha+\beta+\gamma_Z)\moyenne{xz}&=&\alpha\moyenne{z}+k_Z{\bar b_Z}\moyenne{x}\moyenne{y}.
\\
(\alpha+\beta+\gamma_Y+\gamma_Z)\moyenne{xyz}&=&\alpha\moyenne{yz}+k_Y{\bar b_Y}\moyenne{xz} \nonumber \\
&+&k_Z{\bar b_Z}\moyenne{x}\moyenne{y^2}.
\end{eqnarray}
 The set of equation being close, we obtain the following steady-state coefficient of variation squared for $Z$
\begin{eqnarray}\label{eqNoise}
& &CV_Z^2
=\frac{1+{\bar b_Z}}{\moyenne{z}}\\
& &+\frac{(1+{\bar b_Y})}{\moyenne{y}}
\left\{
1-\moyenne{x}\frac{\gamma_Y(\alpha+\beta+\gamma_Z)}{(\alpha+\gamma_Z)(\gamma_Y+\gamma_Z)+\beta\gamma_Z}
\right\}
.\nonumber
\end{eqnarray}
In the limit $\beta\rightarrow0$ we have $\moyenne{x}=1$, leading to
\begin{eqnarray}
CV_Z^2|_{\beta\rightarrow0}
=
\frac{1+{\bar b_Z}}{\moyenne{z}}+
\frac{\gamma_Z}{\gamma_Y+\gamma_Z}\frac{1+{\bar b_Y}}{\moyenne{y}}.
\end{eqnarray}
Note that the latter result is similar to \eqref{EQ_21} presented earlier in section \ref{mod2}.
It follows that the effect of the upstream regulation onto the noise in $Z$ downstream production can be quantified as:
\begin{eqnarray}\label{eqDeltaNoise}
CV_Z^2-CV_Z^2|_{\beta\rightarrow0}&=&
\frac{1+{\bar b_Y}}{\moyenne{y}}\frac{\gamma_Y}{\gamma_Y+\gamma_Z}(1-\moyenne{x})
\nonumber\\
&\times&\frac{\gamma_Z(\alpha+\beta+\gamma_Y+\gamma_Z)}{(\alpha+\gamma_Z)(\gamma_Y+\gamma_Z)+\beta\gamma_Z}.
\end{eqnarray}
Note that the difference $CV_Z^2-CV_Z^2|_{\beta\rightarrow0}$ is always positive. Even if upstream noise has no direct effect on the distribution of $Y$ (and no effect on the mean number $\moyenne{z}$), this result shows that the feedforward motif leads to an increase of the noise $CV_Z^2$ in further downstream products. 

\begin{figure}[thpb]
\centering
  \includegraphics[width=0.9\linewidth]{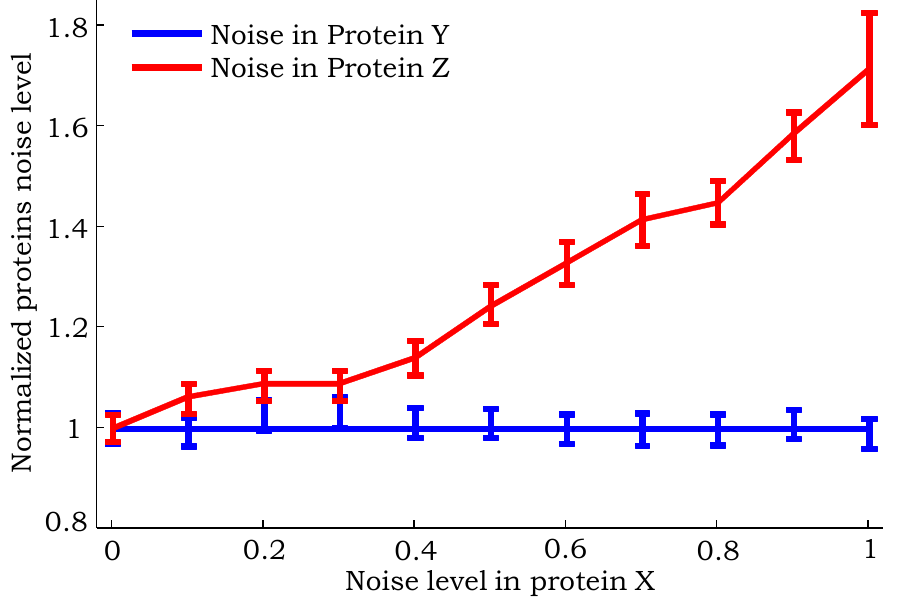}\\
\caption{\textbf{Noise levels in the molecular counts of $Y$ and $Z$ as a function of noise in the upstream regulator $X$ for circuit shown in Fig. 1.} Noise in the levels of $X$ is changed by varying $k_X$ and $\bar b_X$ simultaneously but keeping the mean level $\moyenne{x}=100$ constant. Noise is quantified by the steady-state coefficient of variation squared and normalized with the respective noise levels for $CV^2_X=0$. All other parameters are fixed to $\gamma_X=\gamma_Y=\gamma_Z=1\ hr^{-1}$, $k_Y=10\ hr^{-1}$, $k_Z=20\ hr^{-1}$, and $\bar b_Y=\bar b_Z=1$.
}
\label{simulations}
\end{figure}

The above results were illustrated for an upstream regulator $X$ modeled as a random switch, since exact analytical solutions for statistical moments were available. However, these results also hold qualitatively for a bursty birth-death process. In figure \ref{simulations} we present noise measurements for the feedforward motif illustrated on figure 1, where $x(t)$ is a bursty birth-death process. These results are obtained by averaging a large number of Monte Carlo simulations performed using the Stochastic Simulation Algorithm \cite{gil01}.
Results confirm that $CV_Y^2$ is independent of the noise in $X$. Moreover it clearly shows an increase in downstream product noise $CV_Z^2$, with increasing noise in $X$.

\section{Summary}
Interesting features and new challenges are emerging from the study of biological systems regulated by upstream chemical processes and feedforward genetic motif. Our analysis started with a simple model describing a bursty production and single degradation of molecules $Y$. As expected, when the creation process is regulated by an upstream process, a clear signature of the input noise $X$ is seen in both first \eqref{eqmeanmod2} and second order moments \eqref{EQ_21} of the $Y$-distribution. However, when upstream regulation comes to affect both creation and degradation process, forming a feedforward circuit, we were able to show that all $x$-$y$ correlations vanishes. Thus, the feedforward regulation completely buffers $Y$ from random fluctuations in the upstream regulator and  $x$ appears as a "hidden" variable. We show that a first signature of the existence of $x$ can be found in dynamical quantities. The autocorrelation function was calculated exactly for a feedforward circuit regulated by a simple switch \eqref{eqRsumary}. Our results show dependence in the switch activity and a systematic increase of correlation time scales \eqref{eqRratio}. Interestingly, an indication of noise in upstream regulatory processes can be found in the distribution of further downstream products \eqref{eqNoise}. Here we have shown that the feedforward motif leads to an increase of noise in downstream product \eqref{eqDeltaNoise} leaving however the mean count of molecules $\moyenne{z}$ invariant. In addition, identical observations were confirmed by Monte Carlo simulations (see figure \ref{simulations}) of the more sophisticated model with bursty creation and single degradation of $X$.

The incoherent feedforward circuit considered here is highly simplified, and in reality these systems often involve often biochemical species. For example, instead of $X$ directly activating the production of $Y$, it activates it via an intermediate specie \cite{alo07}. One way to incorporate such intermediate species is by introducing time delays. In future work we will investigate stochastic dynamic of circuits where the regulatory effects of $X$ on $Y$ are time delayed. The delays could come in either activation or degradation, and the delay itself could be a random variable.

\section*{ACKNOWLEDGMENT}
\thanks{AS is supported by the National Science Foundation Grant DMS-1312926.}\\

\bibliographystyle{IEEEtran}
\bibliography{references,thesis}


\end{document}